\documentclass[11pt]{article}
\usepackage{graphicx}
\usepackage[margin=1.25in]{geometry}
\usepackage[usenames,dvipsnames]{color}
\usepackage{url}
\usepackage[colorlinks = true,
            linkcolor = blue,
            urlcolor  = blue,
            citecolor = blue,
            anchorcolor = blue]{hyperref}
\usepackage{xspace}
\usepackage{placeins}

\newcommand\pubdate{\today}

\def\Title#1{\begin{center} {\LARGE #1 } \end{center}}
\def\Author#1{\begin{center}{ \sc #1} \end{center}}
\def\Address#1{\begin{center}{ \it #1} \end{center}}

\newcommand\pubblock{\rightline{\begin{tabular}{l} 
         \pubdate \end{tabular}}}
\newenvironment{Abstract}{\begin{quotation} \begin{center}
                       ABSTRACT
     \end{center}\bigskip  }{\end{quotation}}

\newcommand*{\ttbar}{\ensuremath{t\bar{t}}\xspace}

\def\beq{\begin{equation}}
\def\eeq#1{\label{#1}\end{equation}}
\def\eeqn{\end{equation}}

\newenvironment{Eqnarray}%
   {\arraycolsep 0.14em\begin{eqnarray}}{\end{eqnarray}}
\def\beqa{\begin{Eqnarray}}
\def\eeqa#1{\label{#1}\end{Eqnarray}}
\def\eeqan{\end{Eqnarray}}

\let\bar=\overbar

\def\lsim{\mathrel{\raise.3ex\hbox{$<$\kern-.75em\lower1ex\hbox{$\sim$}}}}
\def\gsim{\mathrel{\raise.3ex\hbox{$>$\kern-.75em\lower1ex\hbox{$\sim$}}}}

\def\del{\partial}
\def\Dslash{\not{\hbox{\kern-4pt $D$}}}
\def\dslash{\not{\hbox{\kern-2pt $\del$}}}
\def\pslash{\not{\hbox{\kern-2pt $p$}}}
\def\ETmiss{\not{\hbox{\kern-4pt $E$}}_T}

\def\Dlr{\mathrel{\raise1.5ex\hbox{$\leftrightarrow$\kern-1em\lower1.5ex\hbox{$D$}}}}

\def\MSB{{\bar{M \kern -2pt S}}}
\def\msb{{\bar{\scriptsize M \kern -1pt S}}}

\def\drb{{\bar{\scriptsize D \kern -1pt R}}}

\def\MeV{{\rm MeV}\xspace}
\def\GeV{{\rm GeV}\xspace}
\def\TeV{{\rm TeV}\xspace}

\newcommand\snowmass{\begin{center}\rule[-0.2in]{\hsize}{0.01in}\\\rule{\hsize}{0.01in}\\
\vskip 0.1in Submitted to the  Proceedings of the US Community Study\\ 
on the Future of Particle Physics (Snowmass 2021)\\ 
\rule{\hsize}{0.01in}\\\rule[+0.2in]{\hsize}{0.01in} \end{center}}

\begin{document}
\pubblock

\Title{Dependence of the top-quark mass measured in top-quark pair production on the parton distribution functions at the LHC and future colliders}

\bigskip 

\Author{Jason Gombas, Jarrett Fein, Sara Sawford, Reinhard Schwienhorst}

\medskip

\Address{Michigan State University, East Lansing, MI 48824, USA}

\medskip

 \begin{Abstract}
\noindent The dependence of the top-quark mass measurement in top-quark production on the parton distribution functions is explored through the differential distribution of the invariant mass of the top-antitop system in top-quark pair production at hadron colliders. Several different proton-proton collider options are considered: 8\ \TeV, 13\ \TeV, 13.6\ \TeV, 14\ \TeV, and 100\ \TeV. The top-quark mass is obtained from a chi-squared fit to the invariant mass of the top-antitop-quark pair. The parton distribution function uncertainties are used in the chi-square evaluation. The uncertainties are reduced through a fit to the longitudinal momentum of the top-antitop system. The top-quark mass uncertainty due to parton distribution functions is found to be reduced by 20\%.  
\end{Abstract}

\snowmass

\def\thefootnote{\fnshttps://www.overleaf.com/project/6203012c02270a6033bd4918ymbol{footnote}}
\setcounter{footnote}{0}
\clearpage

\section{Introduction}
\label{sec:intro}
The top quark mass is one of the most important fundamental parameters~\cite{ParticleDataGroup:2020ssz}. It is one of the key ingredients to global electroweak fits~\cite{Haller:2018nnx}, and a precision measurement of the top quark mass also determines if we live in a stable, unstable, or meta-stable universe~\cite{Degrassi:2012ry}. Future measurements of the top quark mass are limited by theoretical uncertainties more than experimental ones~\cite{Hoang:2020iah}. 

At hadron colliders, the most precise measurements of the top-quark mass reconstruct the top quark from its decay products and then extract a mass through a comparison of data and simulation~\cite{ATLAS:2018fwq,CMS:2015lbj,CMS:2018tye}. These measurement of the "Monte Carlo" mass assume that the top-quark mass parameter in the MC shower packages Pythia~\cite{Sjostrand:2014zea} or Herwig~\cite{Bellm:2015jjp} is closely related to the top-quark pole mass~\cite{Hoang:2020iah}.

Measurement of the top-quark pole mass can be made by unfolding differential top-quark pair (\ttbar) event distributions to the parton level~\cite{ATLAS:2017dhr,CMS:2019esx} and then comparing those distributions directly to precise QCD calculations~\cite{Guzzi:2014wia,Czakon:2011xx}. This procedure is theoretically well-defined, but has larger uncertainties. Experimental uncertainties in the unfolding are larger due to the extrapolation to the full phase-space, and the comparisons to theory predictions provide additional theoretical uncertainties. In particular the uncertainty due to the parton-distribution functions (PDFs) is important. For the inclusive cross-section calculation, the PDFs are the dominant uncertainty~\cite{Czakon:2013goa}.

Here we explore a measurement of the top quark mass in \ttbar production at the parton level, focusing on the contribution from parton distribution functions (PDFs) to the uncertainty on the top-quark mass. We fit the top-quark mass to the differential distribution of the invariant mass of the \ttbar system, evaluating a chi-square based on the PDF uncertainty. These uncertainties can be reduced through a fit to the longitudinal momentum ($p_z$) of the top-quark-pair system using ePump~\cite{Kadir:2020yml}. We study proton-proton colliders in various configurations, including the LHC at 8, 13, 13.6, 14~\TeV~\cite{Evans:2008zzb} and the 100~\TeV proton-proton collider~\cite{FCC:2018byv,FCC:2018vvp,CEPCStudyGroup:2018rmc}.

This paper is organized as follows: Section~\ref{sec:setup} explains the simulation setup and assumptions, Section~\ref{sec:topmass} presents the fit of the top-quark mass, Section~\ref{sec:pdfs} the PDF fit, Section~\ref{sec:combined} the combined fit to top-quark mass and PDF, and Section~\ref{sec:conclusion} gives our conclusions.
\FloatBarrier

\section{Simulation setup}
\label{sec:setup}
We generate top-quark pair events with Madgraph~\cite{Alwall:2014hca} for proton-proton colliders at next-to-leading order in QCD (NLO). The PDF set CT18NLO is used~\cite{Hou:2019efy}. We do not include next-to-next-to-leading-order (NNLO) or resummation effects in our calculation~\cite{Czakon:2013goa,Kidonakis:2019yji}. While these are both are important to model the data accuratly at the \ttbar production threshold, we are not concerned with that modeling here and instead focus on the uncertainty limitations, comparisons between CM energies and the impact of the PDFs. We do not use the scale uncertainties in the theory calculation and therefore would not benefit from higher-order corrections and resummation in the context of our study. The factorization and renormalization scales are set to the average transverse mass of the two top quarks (the Madgraph default).

We analyze events at the top-quark level, meaning top quarks do not decay, and there is no final-state radiation. We do not account for experimental acceptance or efficiencies, which is not a problem in this study because experiments unfold their data to the full phase-space. We also do not include statistical uncertainties due to the limited dataset sizes, and instead assume that the systematic uncertainties dominate over statistical ones. Statistical uncertainties were still relevant (though not dominant) at 8\ \TeV~\cite{ATLAS:2017dhr} but are negligible for Run~2, Run~3, HL-LHC and FCC-hh~\cite{CMS:2019esx}.  
The following center-of-mass energies are considered: The LHC~\cite{Evans:2008zzb} with CM energies of 8, 13, 13.6, and 14\ \TeV, as well as the FCC-hh with a CM energy of 100\ \TeV. The parameters are summarized in Table~\ref{tab:colliders}. For the LHC in Run 3 (13.6~\TeV), the differences with respect to HL-LHC (14~\TeV) are small, we therefore do not plot or report numbers for 13.6~\TeV unless it adds information.

\begin{table}
    \centering
    \begin{tabular}{c|c|c}
    Collider     & CM energy [\TeV] & Luminosity [fb$^{-1}$] \\
    LHC 8 \TeV     & 8    & 20 \\
    LHC Run 2      & 13   & 140 \\
    LHC Run 3      & 13.6 & 300 \\
    HL-LHC         & 14   & 1,000 \\
    FCC-hh         & 100 & 20,000
    \end{tabular}
    \caption{Proton-proton colliders considered in this study and their CM energies and integrated luminosities.} 
    \label{tab:colliders}
\end{table}

\FloatBarrier

\section{Fit of the top quark mass}
\label{sec:topmass}

\begin{figure}
    \centering
    \includegraphics[width=0.49\textwidth]{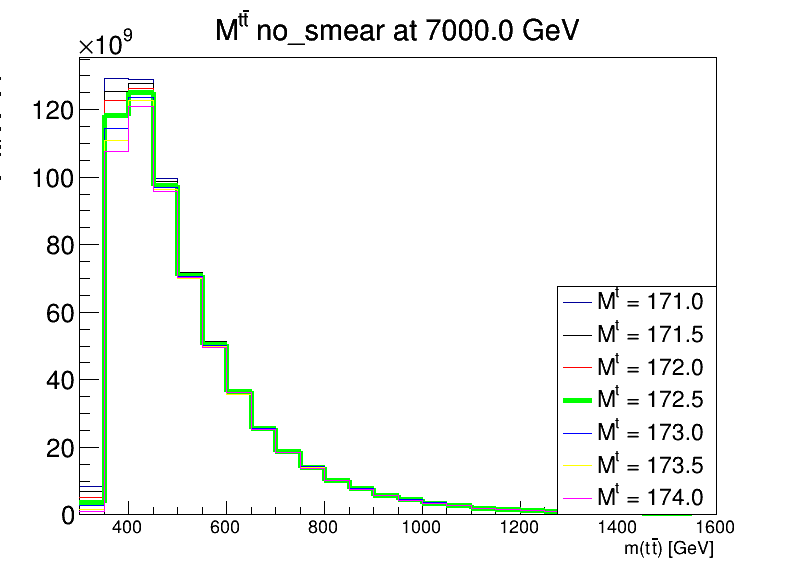}
    \includegraphics[width=0.49\textwidth]{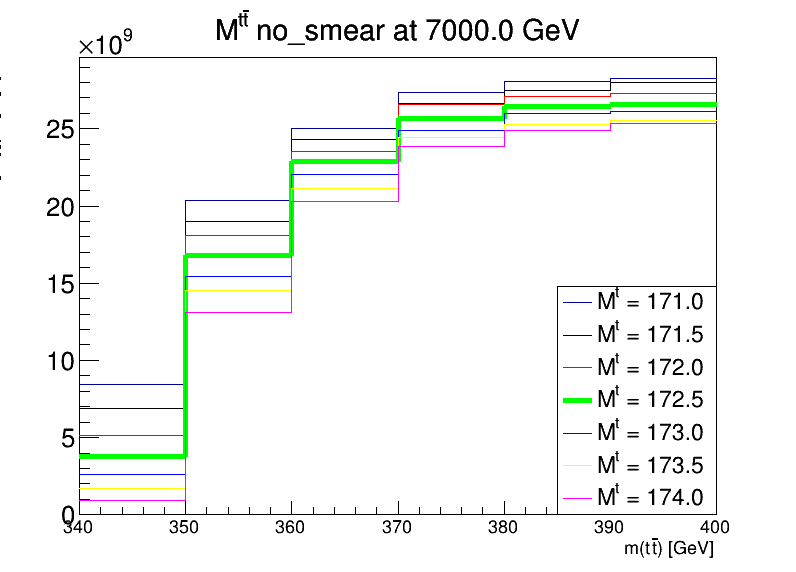}
    \caption{Invariant mass of the \ttbar system for the 14\ \TeV HL-LHC, (left) full distribution and (right) zoom on the threshold region.}
    \label{fig:massdist}
\end{figure}

For \ttbar production, the distribution that is most sensitive to the top-quark mass is the invariant mass of the \ttbar system. This distribution is shown in Figure~\ref{fig:massdist} for the 14~\TeV HL-LHC. The distribution at the threshold is very similar for the other collider options, only the tail to higher masses changes. Most of the sensitivity to the top-quark mass is in the turn-on region around 350~\GeV, which is shown on the right in Figure~\ref{fig:massdist}. The distribution for alternative top-quark masses can also be seen. Each distribution is normalized to the computed cross-section times the integrated luminosity given in Table~\ref{tab:colliders}.

\begin{figure}
    \centering
    \includegraphics[width=0.49\textwidth]{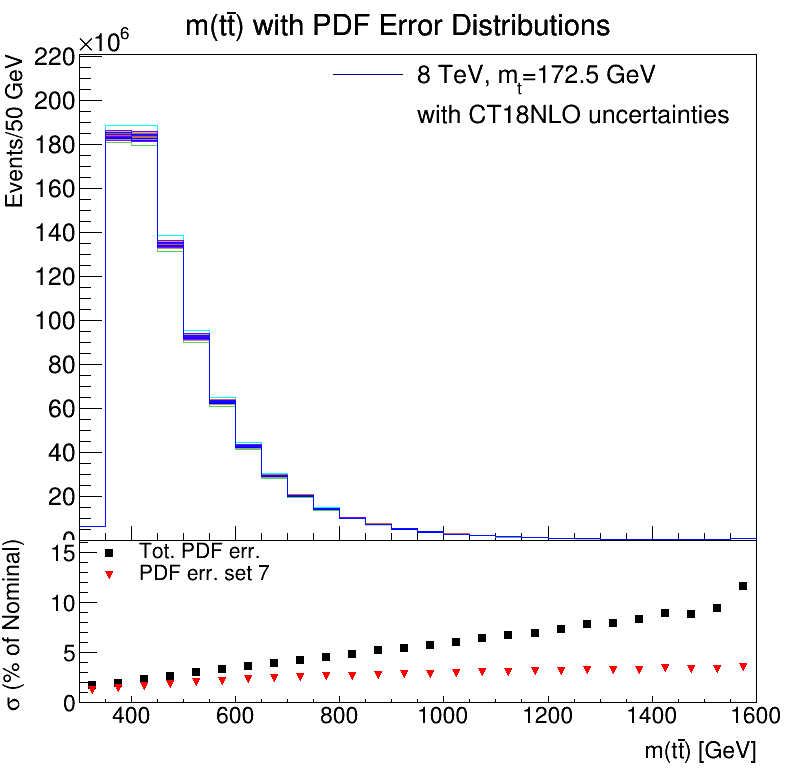}
    \includegraphics[width=0.49\textwidth]{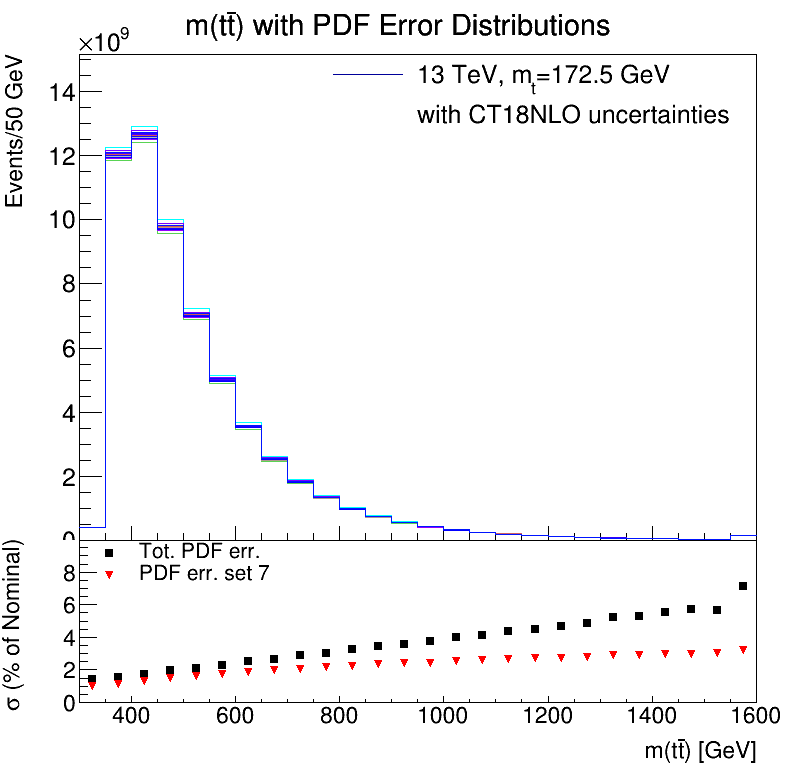}
    \includegraphics[width=0.49\textwidth]{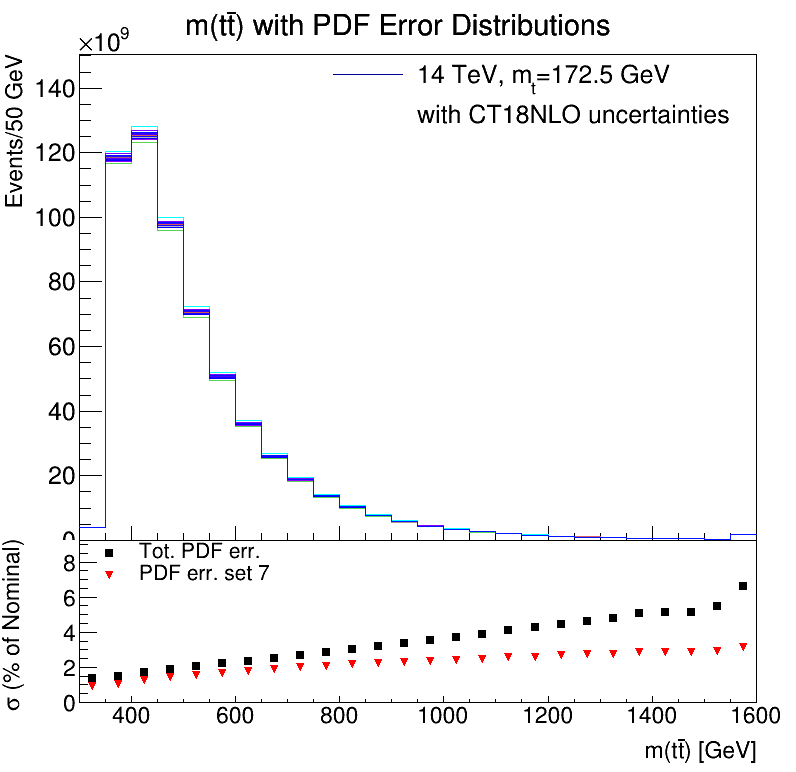}
    \includegraphics[width=0.49\textwidth]{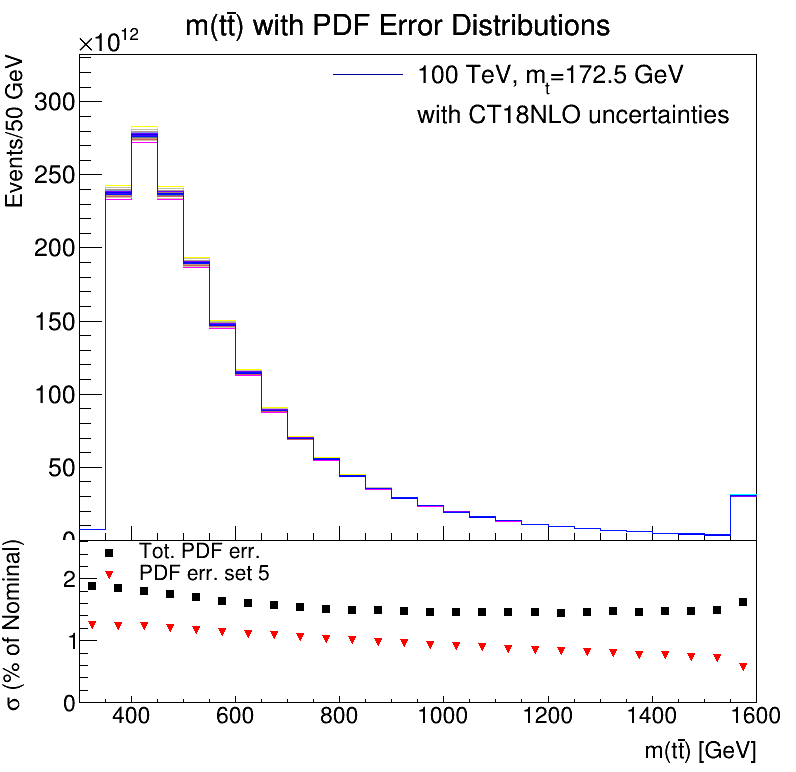}
    \caption{Invariant mass of the \ttbar system for several CM energies of proton-proton colliders. The lower panel in each plot shows the envelope of PDF uncertainties at the 68\% confidence level, as well as the shift from the PDF Eigenvector that contributes the most to the envelope. The last bin includes the overflow.}
    \label{fig:massPDF}
\end{figure}

We evaluate the top-quark mass through a simple chi-squared evaluation. We assume the statistical uncertainty in each bin is negligible, a reasonable assumption for HL-LHC and FCC-hh. These uncertainties are not negligible at the 8~\TeV LHC, but that's not important for this comparison. The distribution at each CM energy shown in Figure~\ref{fig:massPDF} (for a top-quark mass of 172.5~\GeV) is the nominal distribution, with the envelope of PDF uncertainties from CT18NLO providing the uncertainty. For simplicity, we assume that the uncertainties are uncorrelated bin-to-bin in the computation of the chi-square. Figure~\ref{fig:massPDF} shows the top-quark mass distribution and PDF uncertainty at different colliders, including the total PDF uncertainty, and the contribution from the Eigenvector that contributes the most. The PDF uncertainty in the region of low mass, where the sensitivity to the top-quark mass is largest, is small, it rises for larger masses. This is true at all CM energies except at 100\ \TeV, where the PDF uncertainty is more or less constant. This is a result of the higher CM energy of the FCC-hh, which produces top-quark pairs at lower parton momentum fraction $x$ for the incoming partons. Since most of the sensitivity to the top-quark mass is in the first bin of this distribution (see Figure~\ref{fig:massdist}), the PDF uncertainty in this bin determines the sensitivity to the top-quark mass. Here again the 100~\TeV FCC-hh stands out because while it has a small PDF uncertainty for large \ttbar masses, the PDF uncertainty in the first bin (about 2\%) is larger than for all the other configurations (where it is closer to 1.5\%). The 8~\TeV LHC also has a slightly larger uncertainty in this first bin, also close to 2\%.

\begin{figure}
    \centering
    \includegraphics[width=0.89\textwidth]{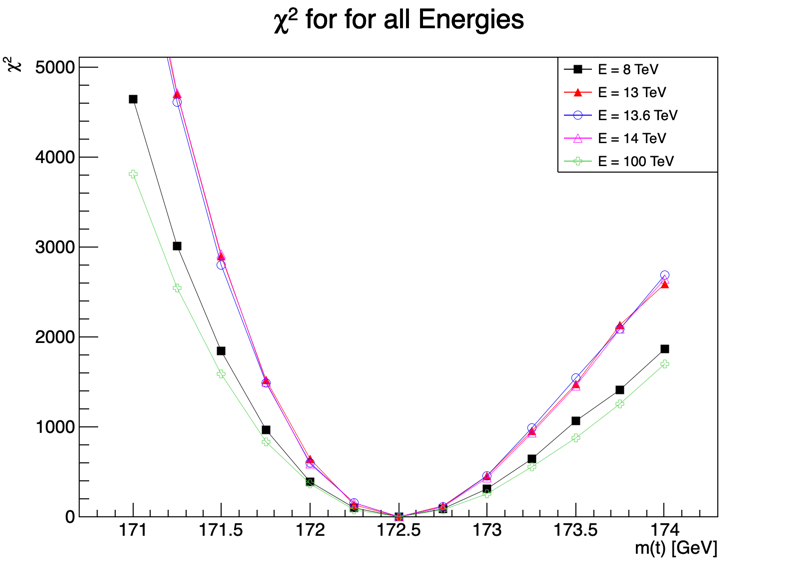}

    \caption{Distribution of the chi-square as a function of top-quark mass for several CM energies of proton-proton colliders, using the envelope of the PDF uncertainties from CT18NLO to compute the chi-square. }
    \label{fig:chi2all}
\end{figure}

The distribution of the \ttbar mass for different top-quark masses provides the pseudo-data for which to evaluate the chi-square. The resulting distribution of chi-square is shown in Figure~\ref{fig:chi2all}. The distribution shows that the 13~\TeV, 13.6~\TeV and `14~\TeV colliders all give roughly the same distribution, as expected since they all have roughly the same PDF uncertainty and the same mass distribution in the first few bins of Figure~\ref{fig:massPDF}. For the 8~\TeV LHC and the 100~\TeV FCC-hh, the PDF uncertainty in the first few bins is slightly larger (see Figure~\ref{fig:massPDF}, which slightly weakens the sensitivity to the top quark mass.

It is possible to also extract a top-quark mass uncertainty from these distributions by taking half of the width when the chi-square is equal to one. The resulting uncertainty is about 25~MeV, which is the contribution to the top-quark mass uncertainty from the PDF uncertainty in the predictions. This uncertainty is small, but we assumed a perfect detector and perfect theory predictions.

\FloatBarrier

\section{PDF fit}
\label{sec:pdfs}

\begin{figure}
    \centering
    \includegraphics[width=0.49\textwidth]{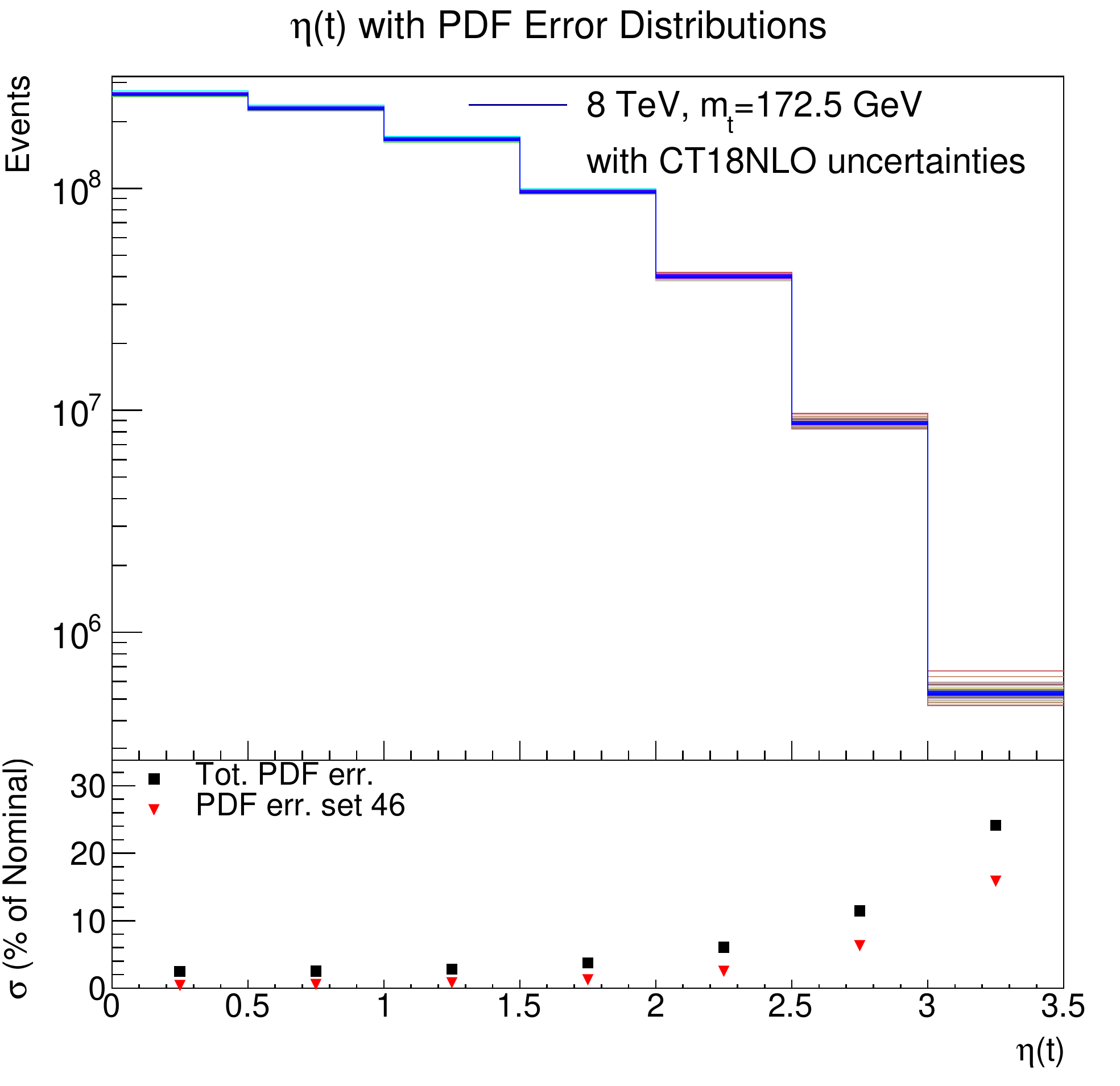}
    \includegraphics[width=0.49\textwidth]{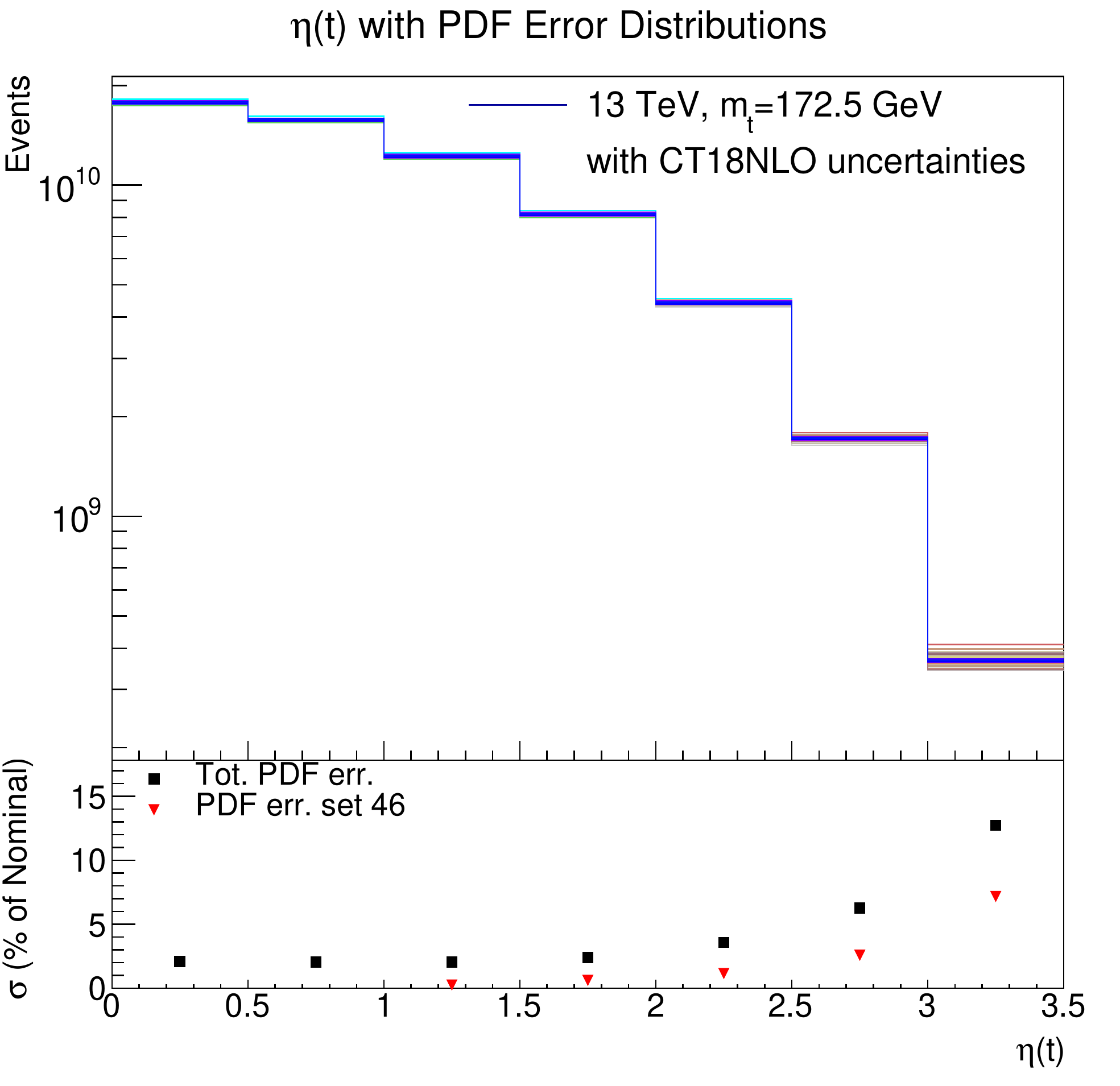}
    \includegraphics[width=0.49\textwidth]{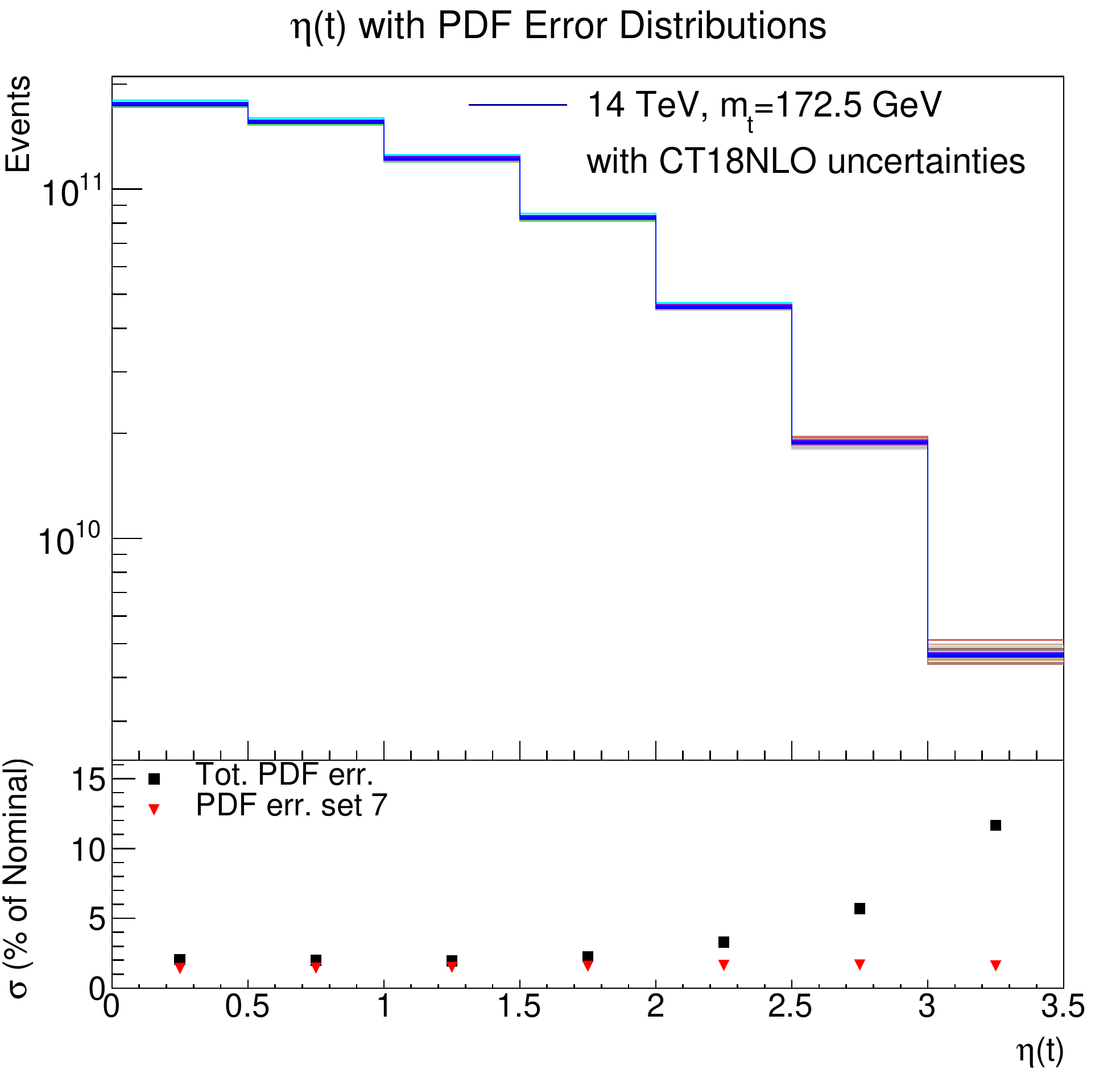}
    \includegraphics[width=0.49\textwidth]{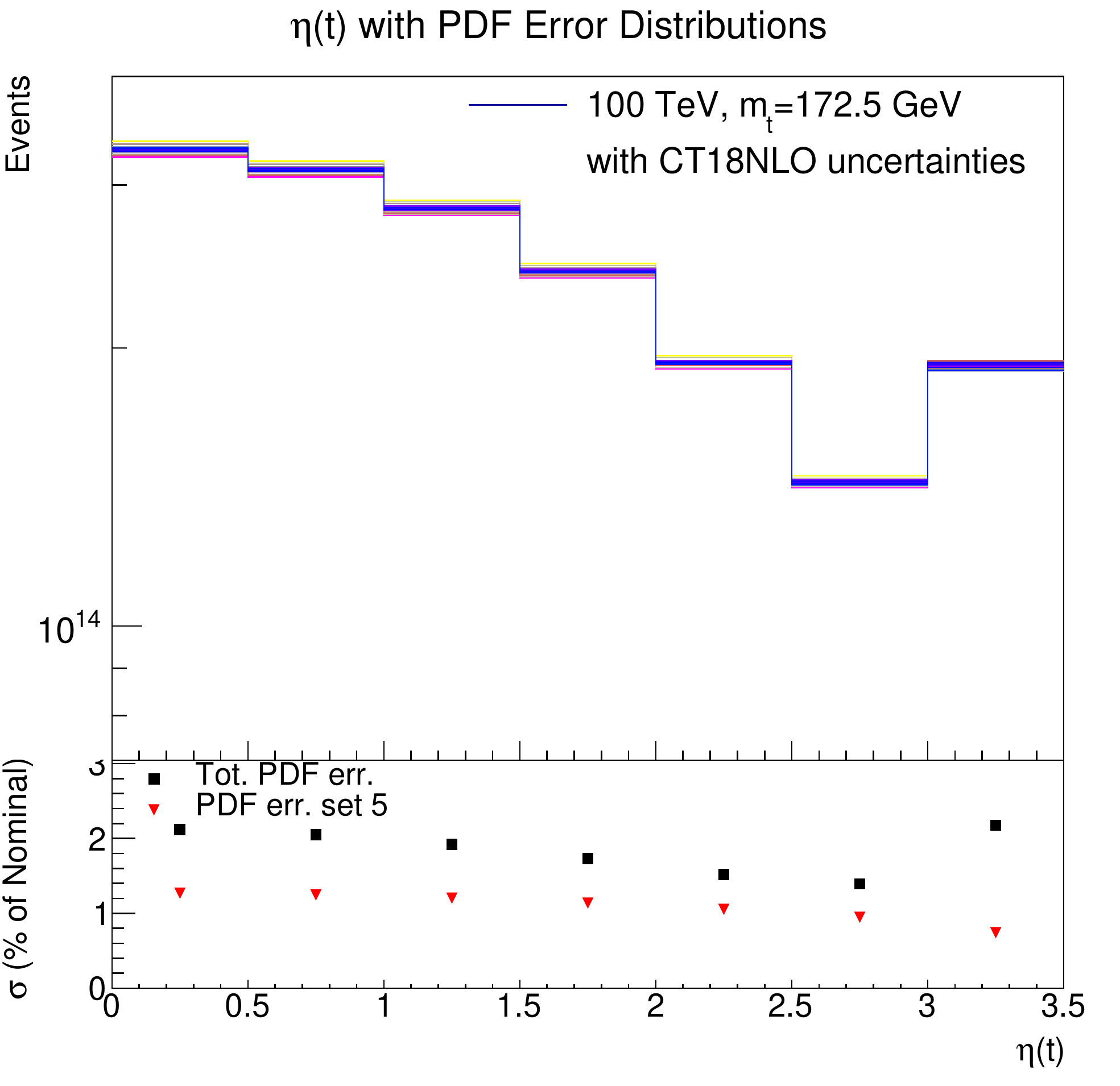}
    \caption{Rapidity of the top quark and the antitop quark for several CM energies of proton-proton colliders. The lower panel in each plot shows the envelope of PDF uncertainties at the 68\% confidence level, as well as the shift from the PDF Eigenvector that contributes the most to the envelope. The last bin includes the overflow.}
    \label{fig:rapidFull}
\end{figure}

The PDF uncertainty for the top-quark mass fit (see Section~\ref{sec:topmass}) can be constrained using the same top-quark dataset as used in the top-quark mass fit. Here we explore how to maximize the constraint from the top-quark pair data on the PDFs. 
Figure~\ref{fig:rapidFull} shows the distribution of the absolute value of the rapidity of the top quark and the anti-top quark (two entries per event, one for the top and one for the antitop). The largest PDF variations can be found for large rapidities ($R>2.5$), but that forward region is very challenging to access experimentally. The typical detector coverage at the LHC ends around 2.5, though upgrades for the HL-LHC should make rapidity up to 3.5 accessible~\cite{CMS-PAS-FTR-18-015}.
The Figure also shows which PDF Eigenvector set contributes the most. It is set 46 at lower CM energies, then set 7 at 14~\TeV, and then set 5 for 100~\TeV.

\begin{figure}
    \centering
    \includegraphics[width=0.49\textwidth]{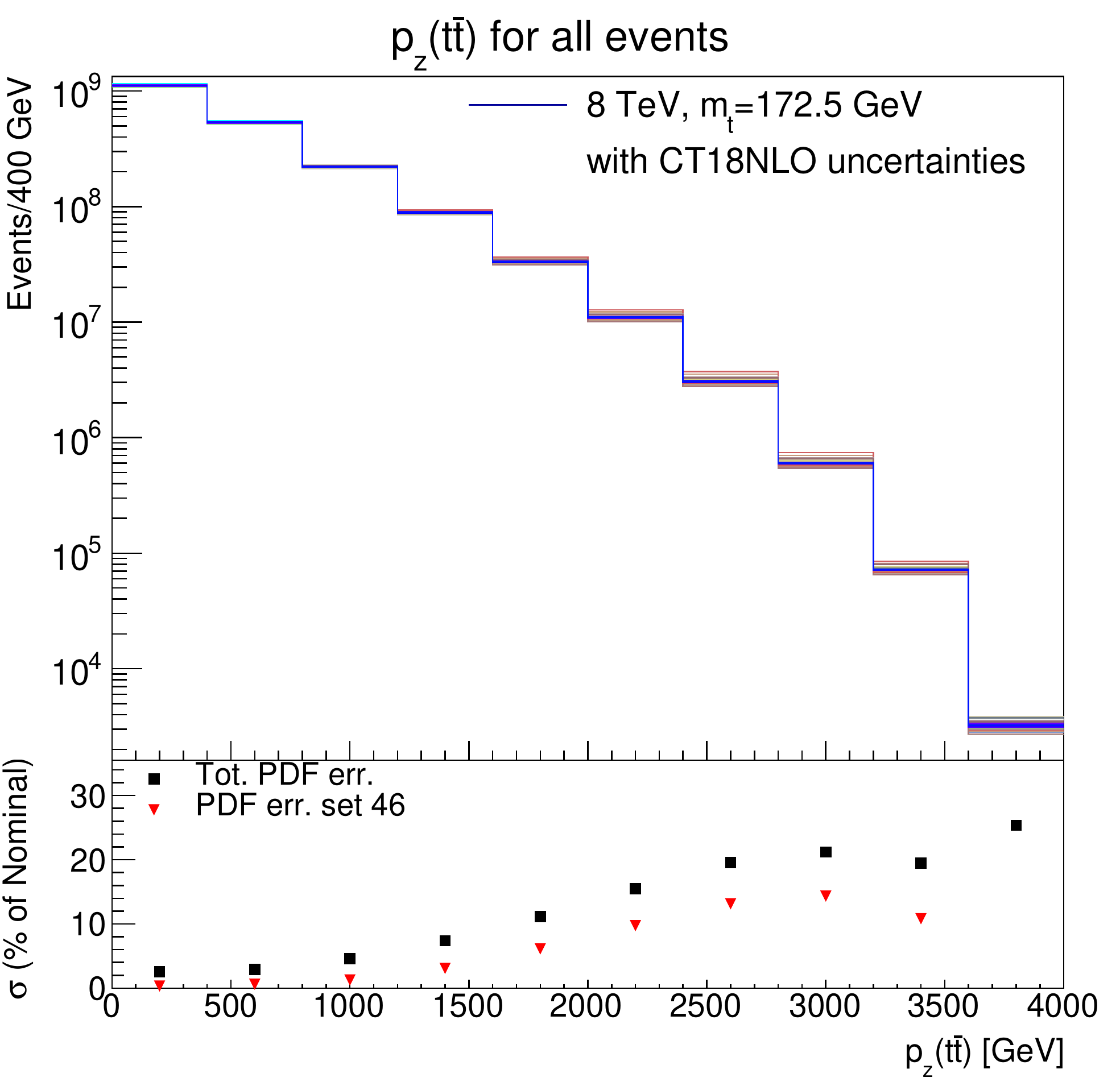}
    \includegraphics[width=0.49\textwidth]{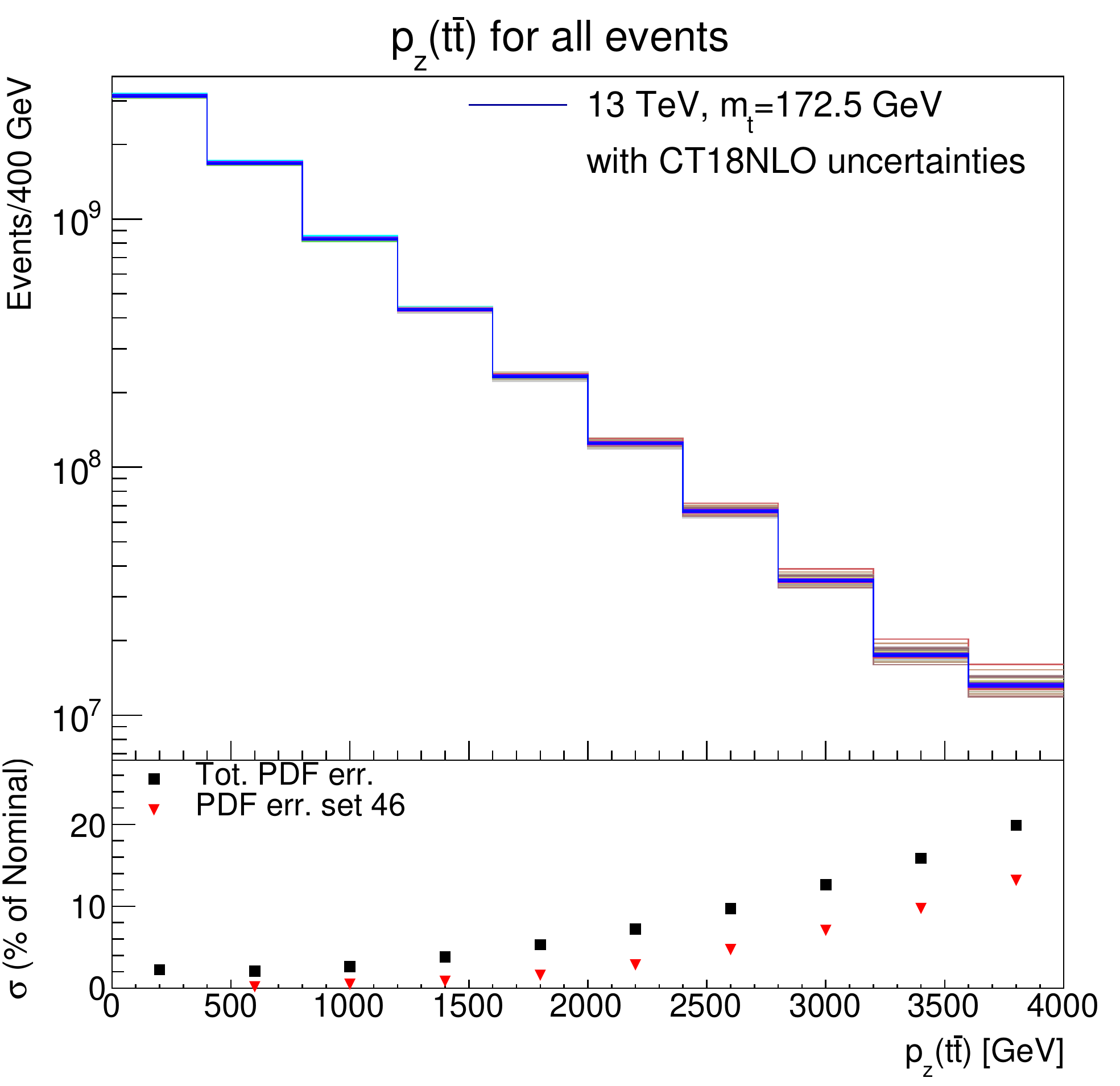}
    \includegraphics[width=0.49\textwidth]{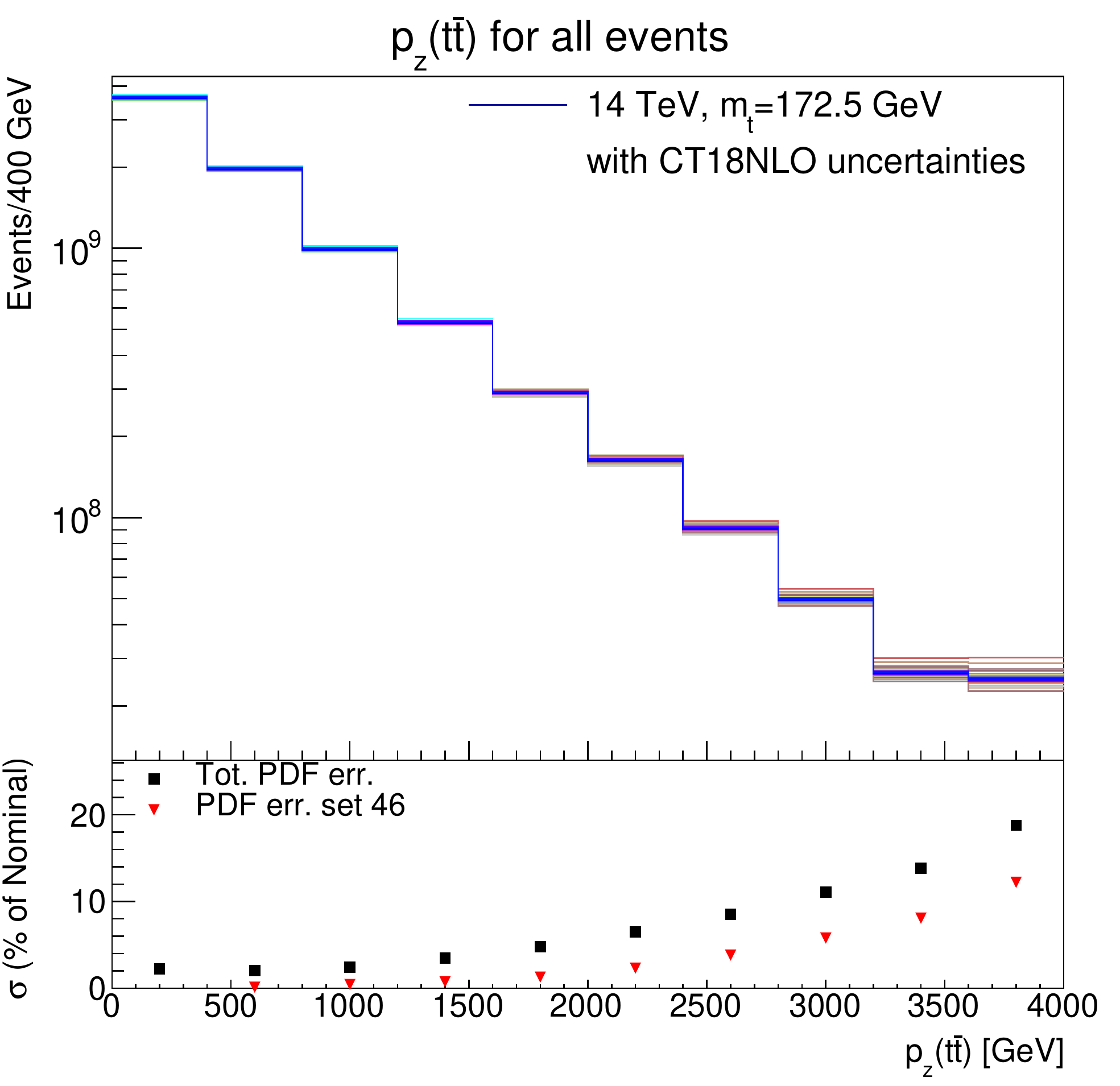}
    \includegraphics[width=0.49\textwidth]{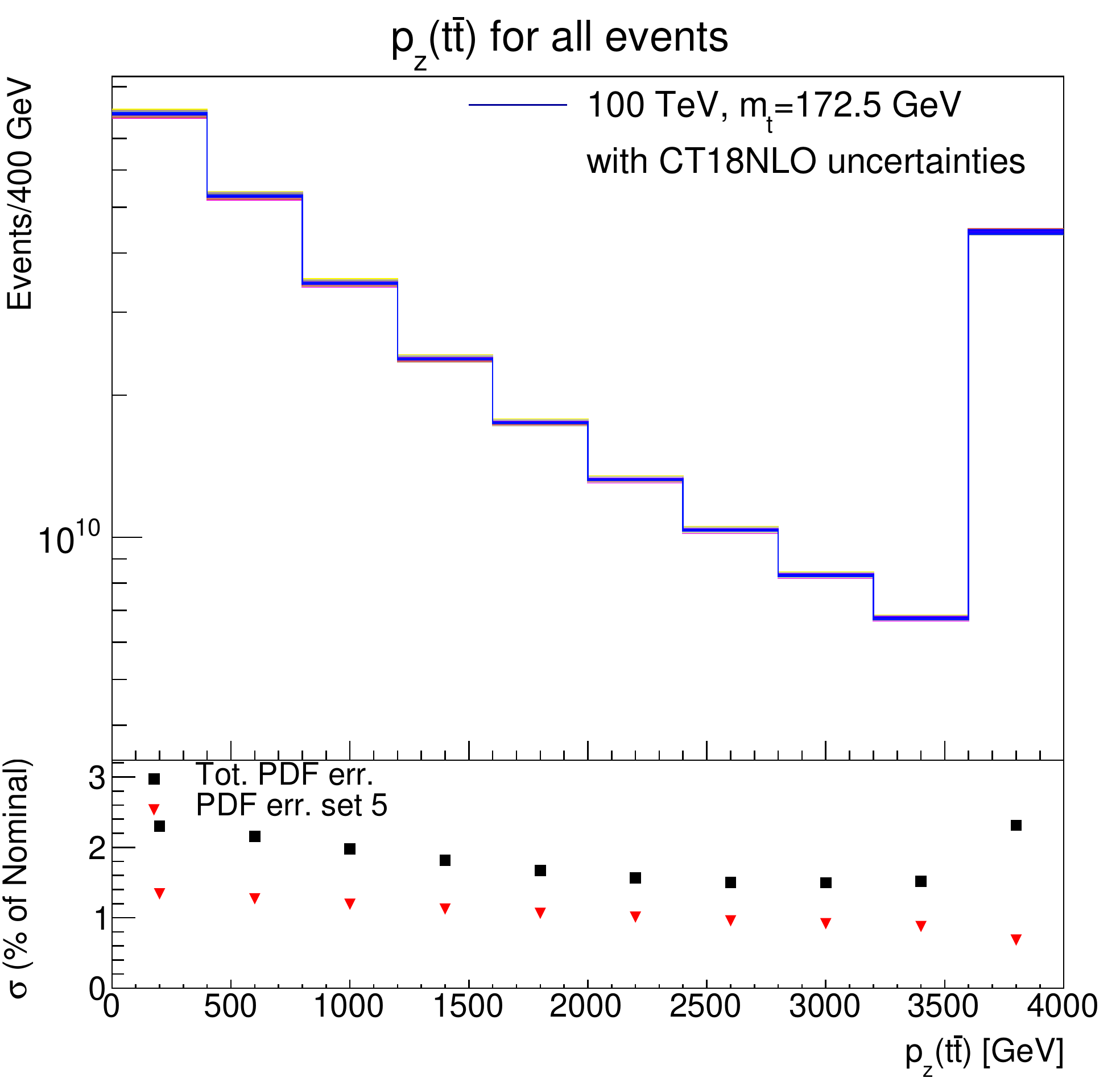}
    \caption{Longitudinal momentum of the \ttbar system for several CM energies of proton-proton colliders. The lower panel in each plot shows the envelope of PDF uncertainties at the 68\% confidence level, as well as the shift from the PDF Eigenvector that contributes the most to the envelope. The last bin includes the overflow.}
    \label{fig:pzFull}
\end{figure}

As an alternative, consider the longitudinal momentum $p_z$ of the \ttbar system. This variable is shown in Figure~\ref{fig:pzFull}. The Eigenvector sets that contribute the most are 46 for the lower CM energies (same as for the rapidity distribution except at 14~\TeV), and set 5 for the 100~\TeV collider. Note that these are not the PDFs that are dominating the top-quark mass, where it is set 7. Nevertheless, the PDF fit constrains several different PDF Eigenvectors, thus there will be an overall reduction.

\begin{figure}
    \centering
    \includegraphics[width=0.49\textwidth]{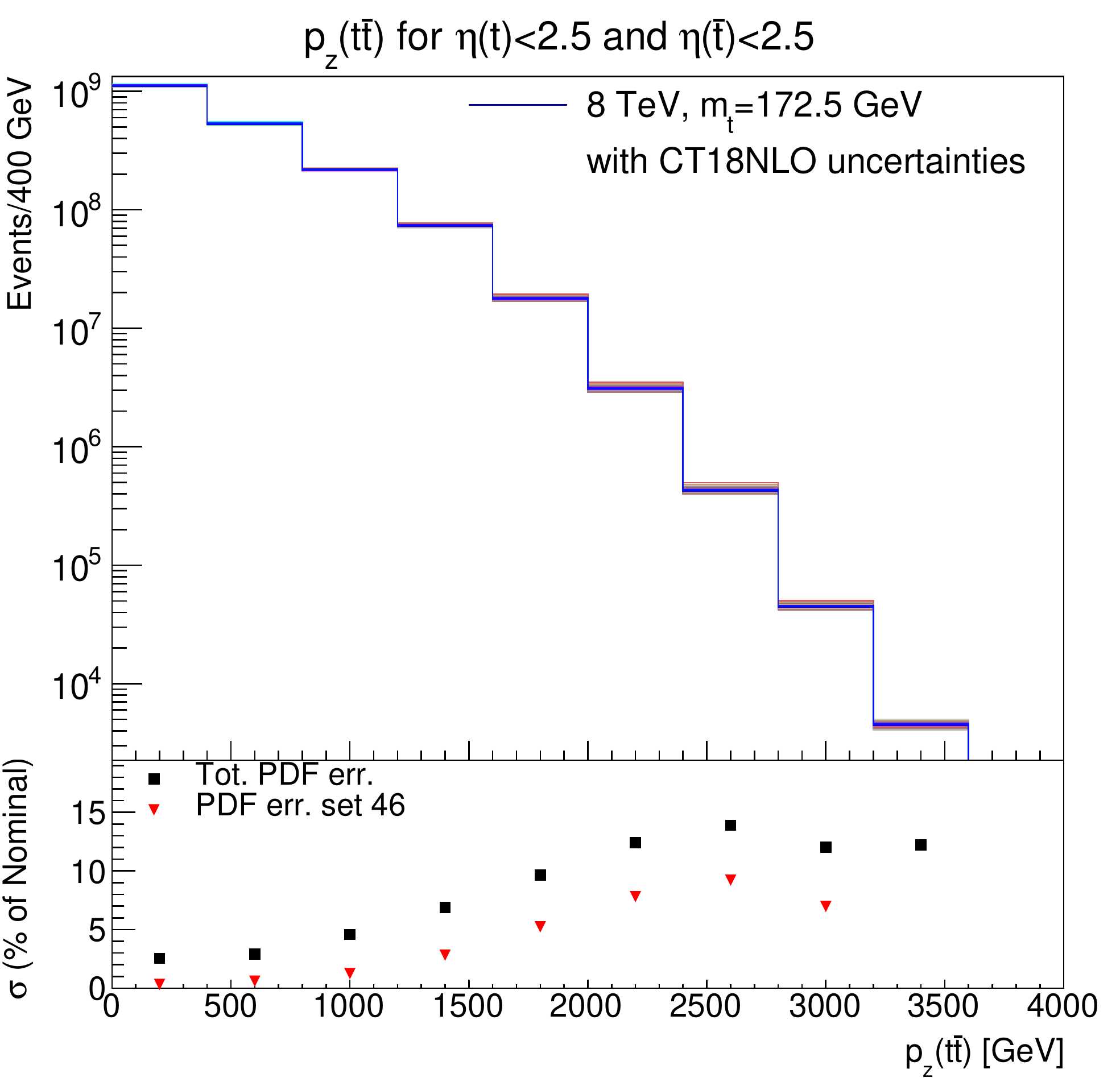}
    \includegraphics[width=0.49\textwidth]{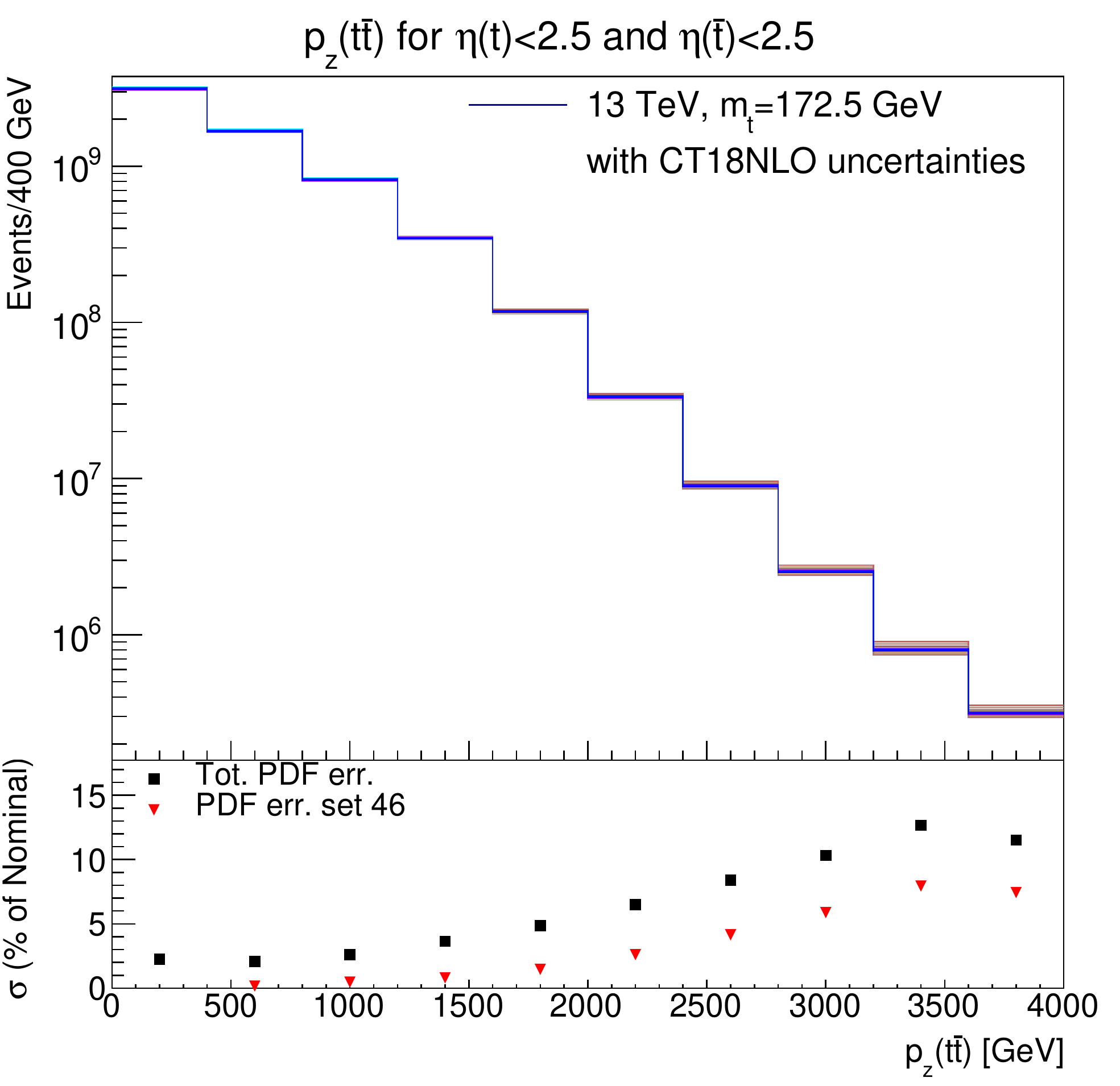}
    \includegraphics[width=0.49\textwidth]{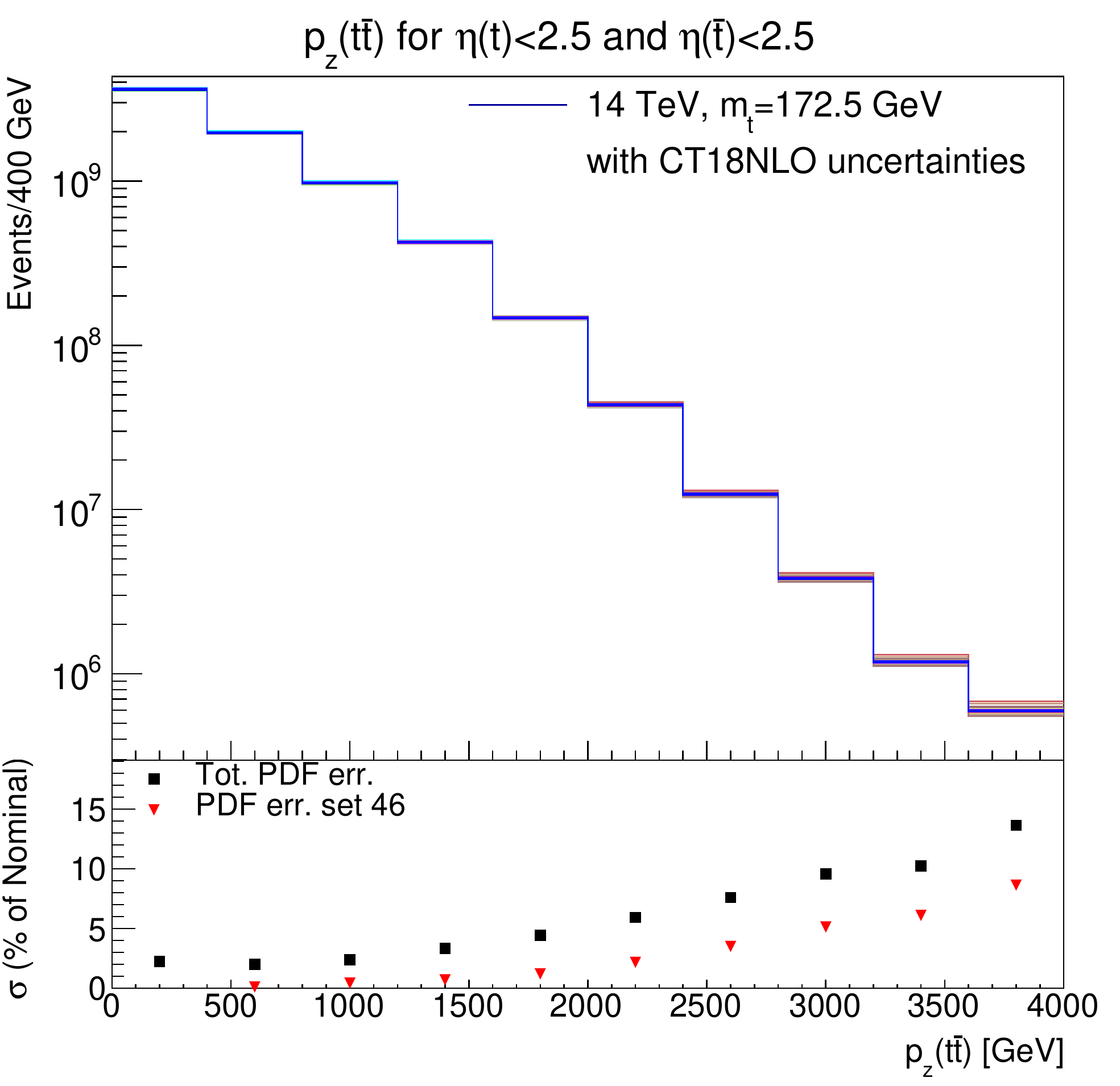}
    \includegraphics[width=0.49\textwidth]{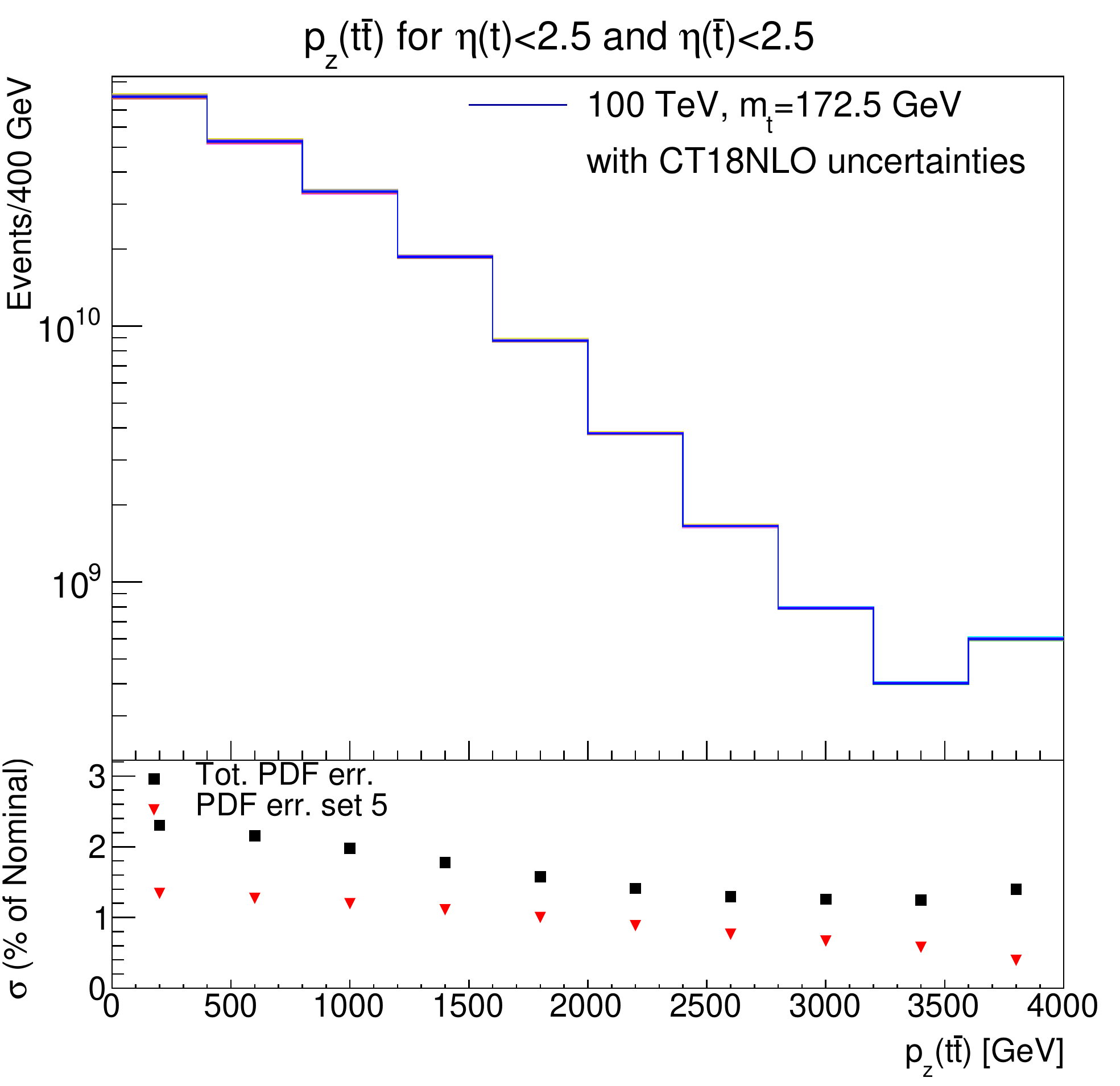}
    \caption{Longitudinal momentum of the \ttbar system for several CM energies of proton-proton colliders, for events where the rapidities of both the top quark and the antitop quark are less than 2.5 ($|R|<2.5$). The lower panel in each plot shows the envelope of PDF uncertainties at the 68\% confidence level, as well as the shift from the PDF Eigenvector that contributes the most to the envelope. The last bin includes the overflow.}
    \label{fig:pzsmallRap}
\end{figure}

The PDF uncertainties are large for the largest $p_z$ values, over 20\% for the lower CM energies. The PDF uncertainties are small for the 100~\TeV collider as explained before.
The large PDF uncertainties for the lower CM energies only appear at the highest $p_z$ values of several \TeV. It is therefore interesting to see if these events will still end inside the detector volume. Figure~\ref{fig:pzsmallRap} shows the $p_z$ distribution of the \ttbar system, requiring that both the top quark and the antitop quark are central ($|R|<2.5$). The distributions have fewer events at the highest $p_z$ values, and the PDF uncertainties for large $p_z$ are reduced to around 15\% for the lower CM energies. But thanks to the large integrated luminosities expected, especially at the HL-LHC, there are still over a million top-quark pairs produced with a $p_z>3000$~GeV. It should be possible to reconstruct these to constrain PDFs tightly.

\begin{figure}
    \centering
    \includegraphics[width=0.49\textwidth]{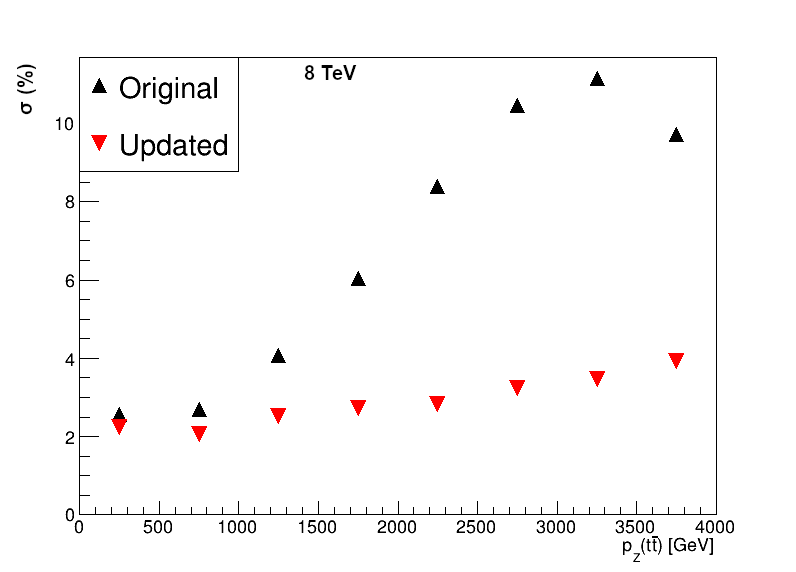}
    \includegraphics[width=0.49\textwidth]{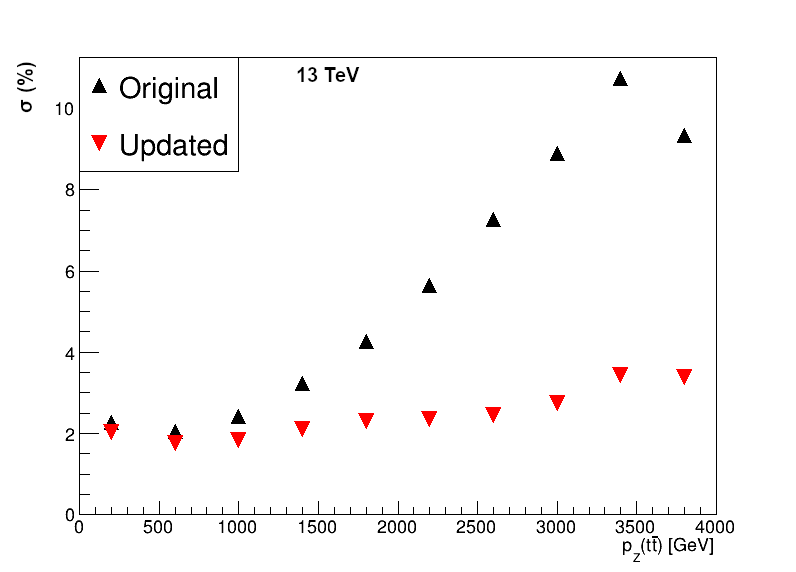}
    \includegraphics[width=0.49\textwidth]{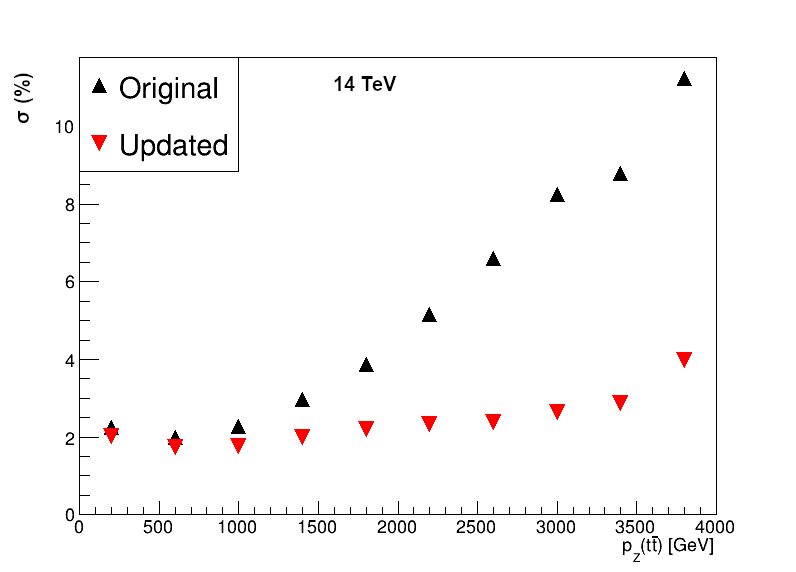}
    \includegraphics[width=0.49\textwidth]{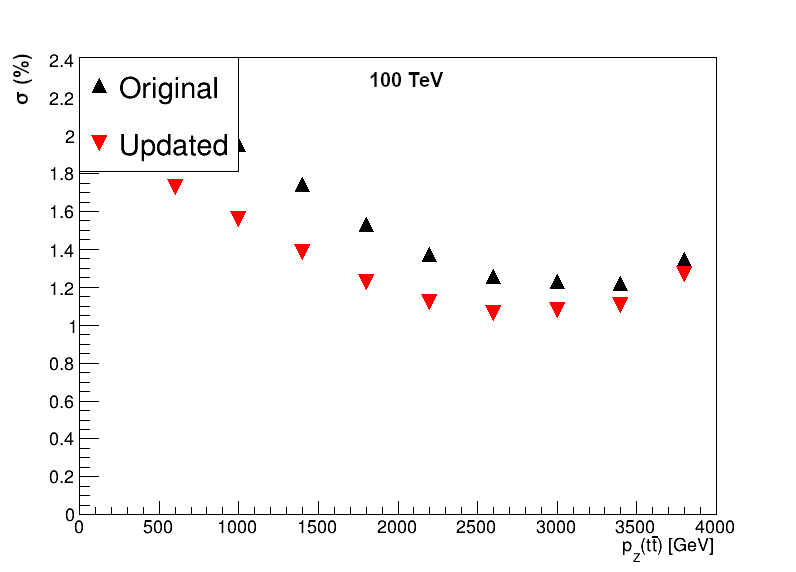}
    \caption{Reduction of the PDF uncertainty (68\%, 1$\sigma$ level) for the distribution of the longitudinal momentum of the \ttbar system for several CM energies of proton-proton colliders, for events where the rapidities of both the top quark and the antitop quark are less than 2.5 ($|R|<2.5$). The vertical axis shows the PDF uncertainty in percent.}
    \label{fig:pzPDF}
\end{figure}

We use the $p_z$ distribution shown in Figure~\ref{fig:pzsmallRap} to constrain the CT18NLO Eigenvectors with ePump. The distribution for 172.5~\GeV forms the prediction, with an uncertainty of 1\% in each bin, uncorrelated across bins. The distribution for 172.5~\GeV also forms the data, so that we do not expect any shifting of the PDFs, only a reduction in uncertainty. The uncertainty that we select is ambitious but should be within reach of the HL-LHC~\cite{ATLAS:2019hxz,CMS:2019esx}, though more precise theory predictions are required in addition to precise measurements. 
The resulting reduction of the uncertainty on the $p_z$ distribution is shown in Figure~\ref{fig:pzPDF}.

The reduction is most pronounced at the high values of $p_z$ where the uncertainties were the largest for the lower CM energies. At 100~\TeV, the uncertainties are not much reduced, but were already below 2\% for most of the distribution.
\FloatBarrier

\section{Top quark mass determination with PDF constraint}
\label{sec:combined}

\begin{figure}
    \centering
    \includegraphics[width=0.49\textwidth]{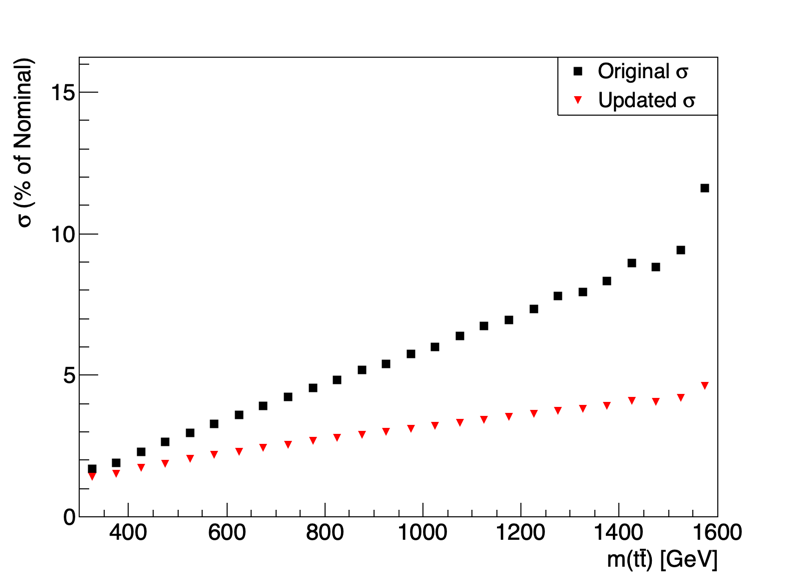}
    \includegraphics[width=0.49\textwidth]{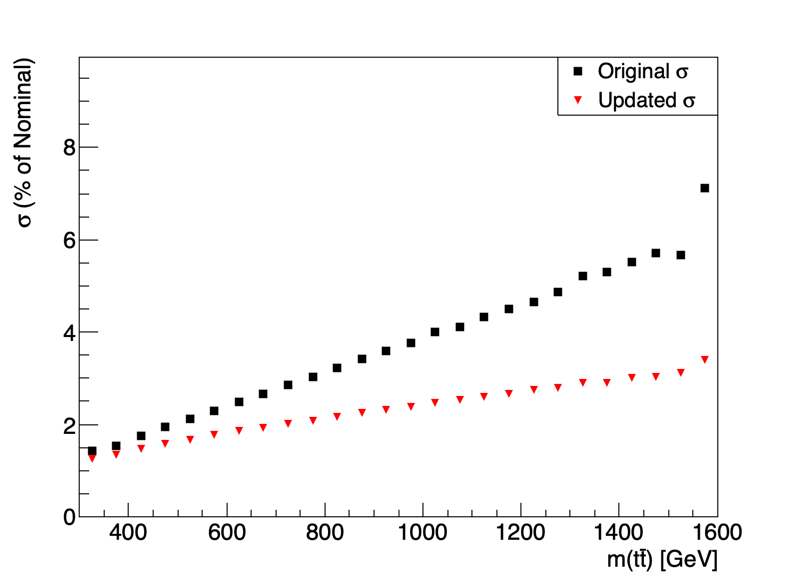}
    \includegraphics[width=0.49\textwidth]{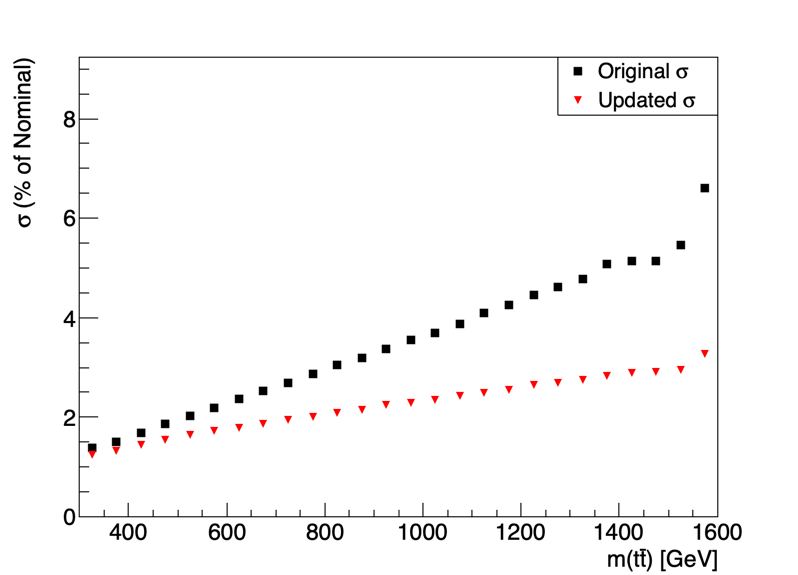}
    \includegraphics[width=0.49\textwidth]{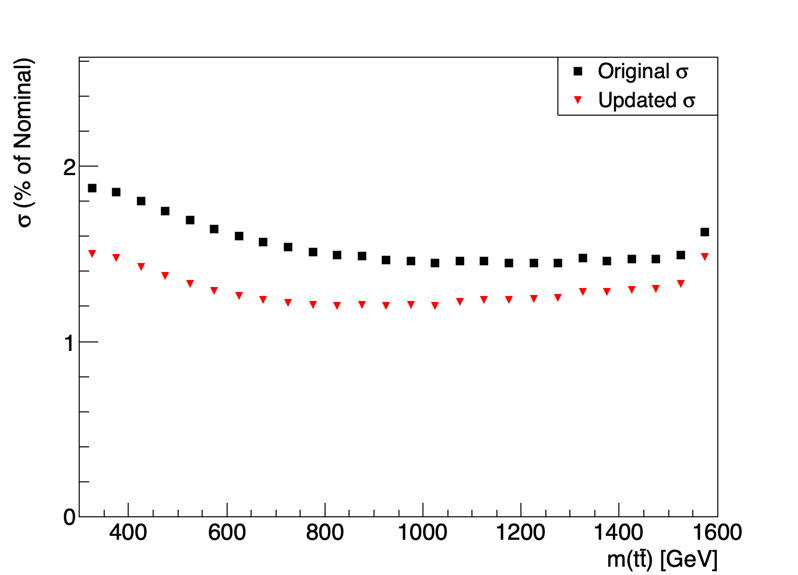}
    \caption{Envelope of the parton distribution function uncertainties before and after constraining them in the fit to the \ttbar system $p_z$. The relative uncertainty (in \%) is shown for the same bins as in Figure~\ref{fig:massPDF}.}
    \label{fig:std}
\end{figure}

We use the updated PDF Eigenvectors, constrained in the fit to the $p_z$ of the \ttbar system, see Section~\ref{sec:pdfs}. 
The PDF uncertainty for the distribution of the mass of the \ttbar system is reduced. Figure~\ref{fig:std} shows the relative uncertainty, the envelope of the PDF uncertainties, without and with constraining them in the ePump fit to the $p_z$ distribution. The uncertainty is calculated as the relative one standard deviation uncertainty, summing the up and down shifts over all Eigenvectors in quadrature and symmetrizing afterwards.  The PDF uncertainties are reduced quite a bit, the reduction achieved in the $p_z$ fit directly translates to the mass distribution.

The constraints are generally stronger for higher \ttbar masses and do not constrain the threshold region as much. The exception to this is the 100~\TeV collider, where even the lowest masses see a reduction in the PDF uncertainty.

\begin{figure}
    \centering
    \includegraphics[width=0.89\textwidth]{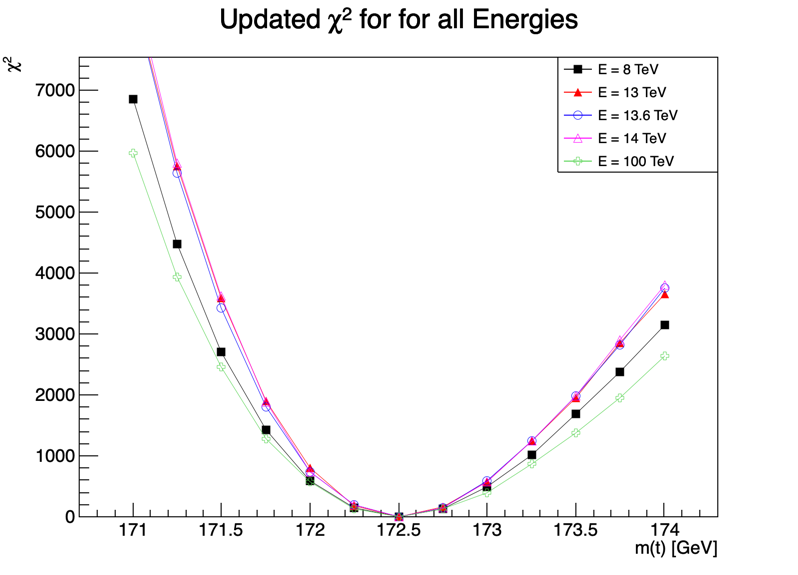}

    \caption{Distribution of the chi-square as a function of top-quark mass for several CM energies of proton-proton colliders, using the envelope of the updated CT18NLO PDF uncertainties. These PDFs were constrained in the fit to the \ttbar $p_z$.}
    \label{fig:chi2updated}
\end{figure}

We use these updated PDFs with their reduced uncertainties and re-compute the chi-square distribution as a function of the top-quark mass from the distribution of the invariant mass of the \ttbar pair, see Section~\ref{sec:topmass}. The result is shown in Figure~\ref{fig:chi2updated} shows the updated chi-square distribution using the updated PDF uncertainties. The distributions for all CM energies are more narrow. The width for a chi-square of one (top-quark mass uncertainty) is about 20~\MeV, so an improvement of 20\% compared to the original PDF uncertainties.

\FloatBarrier

\section{Conclusions}
\label{sec:conclusion}

We have presented a study of the top-quark mass in top-quark pair production at hadron colliders, using the invariant mass of the \ttbar system as the sensitive variable. We study the LHC at 8, 13, 13.6, and 14~\TeV, and the future 100~TeV proton-proton collider. We explore the sensitivity of the top-quark mass fit to parton distribution functions and evaluate the uncertainty on the top-quark mass due to the PDF uncertainty with the CT18NLO PDF set. The PDF uncertainty can be improved by fitting the distribution of the longitudinal momentum $p_z$ of the \ttbar system, which improves the top-quark mass uncertainty by 20\%.
Further improvement of the PDF impact on the top-quark mass measurements in production require additional input from outside top-quark pair production.
\FloatBarrier

\bibliographystyle{JHEP}
\bibliography{myreferences}  


\end{document}